# The Role of High-Speed Rail in Reshaping Chinese County-Level Economic Structures

Mingzhi Xiao[1*], Yuki Takayama[1]

**Abstract:** As high-speed rail (HSR) investment accelerates across China, the question of whether such large-scale infrastructure can promote balanced regional development or exacerbate spatial inequality has become central for policymakers and scholars. This study provides systematic micro-level evidence by analyzing a balanced panel of 353 county-level divisions, including urban districts, county-level cities, and counties, along the Shanghai–Kunming and Xuzhou–Lanzhou HSR corridors from 2008 to 2019. Using a multi-period difference-in-differences (DID) approach, supported by event study and propensity score matching, we quantify the heterogeneous impacts of HSR openings across administrative types and regions, with special attention to the presence of direct HSR station access. The results show that HSR expansion significantly increases secondary and tertiary sector output in urban districts (by 2.77 and 8.71 hundred million RMB) and in county-level cities, particularly in the eastern region. In contrast, counties without HSR stations or with weaker economic foundations experience much smaller gains. Some counties also see a notable contraction in the service sector, which is closely linked to substantial population outflows. Robustness checks confirm the causal interpretation. These findings challenge the prevailing view that HSR fosters uniform growth. Instead, the results reveal that infrastructure-led development can intensify spatial and administrative disparities at the county level. The study underscores the need for integrated and locally tailored policy interventions to ensure that HSR investments contribute to inclusive and sustainable regional development.

# 1 Introduction

Evaluating the economic consequences of major transport infrastructure projects, such as high-speed rail (HSR), is a critical challenge for policymakers seeking to promote balanced regional development and reduce spatial inequality. As governments around the world increasingly invest in HSR systems to enhance connectivity and stimulate growth, understanding who benefits, who does not, and how such investments reshape local economic structures becomes essential for evidence-based policy design. While HSR has the potential to transform regional economies, concerns remain about whether such benefits are distributed evenly across space and sectors.

China provides a unique and analytically rich context to explore these issues. Since the launch of sixth national railway acceleration program in 2007, China has expanded its HSR network at an unprecedented scale and pace. By September 2024, the network had surpassed 46,000 kilometers, aiming to connect nearly all cities with populations over 500,000. The stated goals of this program include reducing intercity travel times to within 1–4 hours and enhancing the cohesion of large urban agglomerations. Beyond improving transport efficiency, the Chinese government views HSR as a strategic tool for promoting regional integration, industrial upgrading, and inclusive development. China's experience thus offers valuable insights into how infrastructure investments interact with varied local conditions to produce differentiated economic outcomes.

---

[1] Institute of Science Tokyo, 2-12-1 W6-9, Ookayama, Meguro-ku, Tokyo 152-8552, Japan
* Corresponding author: xiao.m.1475@m.isct.ac.jp

A substantial body of literature has documented the positive effects of HSR on urban growth, industrial agglomeration, and regional integration (Chen, 2012; Duranton & Turner, 2012; Tierney, 2012). Nevertheless, more recent studies highlight mixed or uneven outcomes, especially when disaggregated to smaller spatial units or differentiated by sector (Hu et al., 2022; Hu & Xu, 2022). For instance, while manufacturing may benefit from improved logistics and labor market integration, service sectors, particularly those reliant on dense population networks or face-to-face interactions, tend to concentrate in urban cores. This "siphoning effect" has been observed in various countries, including China, Japan, and Germany (Hall, 2009; Wang et al., 2024). These patterns raise important questions about whether HSR promotes spatial convergence or, conversely, exacerbates regional inequality.

One underexplored dimension of this debate concerns how HSR affects *sub-city-level divisions*, particularly in countries like China with complex administrative hierarchies. Chinese cities are composed of distinct county-level divisions, *urban districts, counties,* and *county-level cities*, each of which has different institutional roles and levels of connectivity. While urban districts often enjoy direct HSR service, peripheral counties and county-level cities may rely on indirect or multi-transfer access. This variation creates significant heterogeneity in actual accessibility and, potentially, in local economic impacts. Yet, most existing studies focus on city-level averages, obscuring these intra-city differences.

Additionally, few studies have analyzed the role of *station location*, despite its demonstrated importance in other contexts. Evidence from Europe and Japan highlights that the benefits of HSR decline sharply with distance from the station: a phenomenon known as the "last mile" problem (Willigers & Van Wee, 2011; Miwa et al., 2022; Liang et al., 2023). Moreover, station hierarchy and placement influence patterns of population change, investment, and industrial restructuring (Sands, 1993; Wang et al., 2024). Despite China's extensive HSR rollout, little is known about how variation in station access within cities interacts with administrative status and local economic foundations to shape development trajectories.

This study aims to fill that gap. Using a panel dataset of 353 county-level divisions of cities located along two major east–west HSR corridors, we examine how HSR expansion influences industrial development across different sectors (primary, secondary, and tertiary) and administrative types. Particular attention is given to the role of direct HSR station access and regional economic context (eastern, central, and western China). While the empirical focus is China, the findings have broader theoretical and policy implications for countries pursuing infrastructure-led development. By showing how institutional hierarchy and spatial access mediate the effects of HSR, the study contributes to global discussions on transport equity, infrastructure planning, and subnational disparities.

Based on the literature and the specific characteristics of China's spatial governance, we propose the following hypotheses:
- HSR expansion has significant but variable effects on county-level economic development.
- These effects may differ depending on whether a county has direct access to an HSR station.
- Regional context may shape HSR's influence, with possible variation across different parts of China.
- The impacts on peripheral and less-developed counties remain uncertain and may diverge from those in more central or developed areas.

The remainder of this paper is organized as follows. Section 2 reviews the relevant literature on HSR and regional development, highlighting the current gaps in understanding the micro-level effects of HSR at the county division scale. Section 3 details the data sources, sample construction, and empirical strategy, including the difference-in-differences approach and robustness checks used in the analysis. Section 4 presents the main empirical results, examining the overall impact of HSR openings, regional and administrative heterogeneity, and the role of station locations. Section 5 provides an in-depth discussion of the mechanisms behind the observed patterns, evaluates the robustness of the findings, and explores policy implications. Section 6 concludes the paper by summarizing the key findings. Section 7 discusses the broader policy implications derived from the empirical results, with a focus on how infrastructure investment strategies can be optimized to promote balanced regional development. Section 8 addresses the main limitations of the study and suggests directions for future research.

## 2 Literature Review

A substantial body of research has examined how HSR systems shape regional development, industrial restructuring, and urban growth. International evidence shows that HSR can foster agglomeration economies and improve regional connectivity. In some settings, it may also promote the decentralization of economic activity. However, these effects are highly context-dependent. Some studies suggest that HSR helps integrate peripheral regions and supports convergence, while others find that it reinforces growth in already-developed urban cores and increases regional disparities (Sands, 1993; Willigers & Van Wee, 2011; Murakami & Cervero, 2017; Ahlfeldt & Feddersen, 2018; Wang et al., 2024).

In China, HSR has been a central element of national strategies for balanced growth and regional integration. Early research, mostly at the city level, found that HSR openings boost economic output, increase land values, and stimulate investment (Chen, 2012; Jia et al., 2017; Zheng & Kahn, 2013). More recent studies reveal substantial heterogeneity in HSR's impacts across different regions and sectors. The benefits of HSR are particularly pronounced for secondary and tertiary industries in economically developed areas, while agricultural and resource-based regions experience smaller gains (Hu et al., 2022; Qin, 2017; Chi & Han, 2023). These findings highlight the importance of spatial and institutional factors in determining HSR's effects.

Recent research further emphasizes the significance of micro-level heterogeneity and spatial spillovers. For example, Zhang et al. (2023) use county-level panel data and show that HSR openings lead to larger increases in per capita GDP in urban districts than in other areas, with notable differences between direct and spillover effects. Li et al. (2024) use nighttime light data to analyze intra-regional inequality and find that HSR can sometimes reduce disparities in metropolitan areas that have more balanced economic structures. These studies demonstrate the need to consider administrative and institutional diversity when evaluating infrastructure impacts.

Ou et al. (2025) contribute to this discussion with a comprehensive nationwide assessment of both direct and spillover effects of HSR at the county-level divisions, using a difference-in-differences framework. Their findings indicate that HSR stations significantly increase local GDP per capita, with the strongest effects in urban districts and weaker impacts in non-urban counties. They also document that HSR in urban areas can stimulate growth in neighboring non-urban areas but may lead to siphoning effects in adjacent urban districts. Despite these insights, their analysis primarily relies on a binary classification of county-level units as either urban or non-urban. Specifically, Ou et al. classify county-level divisions into "urban" and "non-urban" areas, where urban areas appear to correspond to urban districts, and non-urban areas include both counties and county-level cities.

To clarify the meaning of this classification, it is necessary to understand China's administrative hierarchy. In China, the urban administrative system is organized with prefecture-level cities at the core. Each city is internally divided into several county-level divisions, which are the main units for local governance, economic management, and statistical reporting within the city. These county-level divisions are not the basic units that constitute a city but rather subdivisions that help manage the complex urban system. There are three major types: urban districts, county-level cities, and counties. Urban districts refer to the densely populated and economically active core of the city. County-level cities are urbanized areas with administrative autonomy and significant economic roles, usually located outside the urban core. Counties are generally more rural, less developed, and are primarily responsible for local affairs in peripheral areas.

Ou et al. (2025) group all county-level administrative divisions into just two categories: "urban areas" and "non-urban areas," without distinguishing among urban districts (the core urban area), county-level cities (which are often highly urbanized and economically dynamic), and counties (which tend to be more rural and less developed). While this approach provides a tractable framework for national-level analysis, it may obscure meaningful variation in administrative roles and economic characteristics within cities. For instance, county-level cities can, in many cases, be as urbanized and economically vibrant as urban districts. As a result, important intra-city differences in accessibility, governance, and development potential may not be fully captured in binary classifications.

Another limitation of previous research is the lack of attention to the specific location of HSR stations and actual accessibility within cities. While Ou et al. discuss spatial spillovers, they do not analyze how last-mile connectivity or station siting within urban areas affects local development outcomes. Recent studies in China and abroad have shown that factors such as accessibility and station location can significantly alter the impact of HSR on local economies (Willigers & Van Wee, 2011; Miwa et al., 2022; Liang et al., 2023).

Despite advances in the literature, two major gaps remain. First, most studies use large spatial units such as prefecture-level cities or metropolitan regions, which masks significant heterogeneity within cities. Systematic research at the county-division level is rare, even though many policies and investments in China are implemented at this scale. Second, there is limited focus on how HSR station locations and real accessibility shape economic and demographic outcomes within cities. Although the "last-mile problem" is widely recognized in transportation studies, most Chinese research still treats HSR access as a simple presence-or-absence variable.

The present study aims to address these gaps. By integrating detailed data on both HSR station locations and the administrative hierarchy of county-level divisions, this research provides a more nuanced analysis of how HSR access interacts with local institutional context. Beyond aggregate effects, the study systematically examines outcomes for primary, secondary, and tertiary sectors, and investigates the mechanisms behind divergent development paths, including population outflows and service sector contraction in less-developed areas. The empirical approach combines dynamic event-study analysis, placebo tests, and propensity score matching to strengthen causal inference.

An example from Jinhua City in Zhejiang Province clearly illustrates these issues. Figure 1 shows the administrative divisions of Jinhua City, the layout of HSR routes, and the locations of HSR stations. In this map, blue dots represent the HSR stations on the Shanghai–Kunming HSR corridor analyzed in this study, while red dots indicate HSR stations on a new line that opened in 2024. Table 1 summarizes each division's population, economic profile, and geographic area. The city's urban districts are directly served by centrally located HSR stations, while several county-level cities and counties lack direct access. Residents in these areas often rely on local or bus lines to reach the HSR network, reducing their effective connectivity even when geographically close. This intra-city disparity highlights the importance of considering both administrative status and spatial accessibility when assessing the impacts of HSR.

In summary, there is a clear need for more detailed and spatially sensitive research on HSR effects. By linking sector-specific outcomes to comprehensive data on station locations and administrative divisions, this study offers new empirical evidence on how HSR expansion is reshaping economic structures and spatial inequalities within Chinese cities.

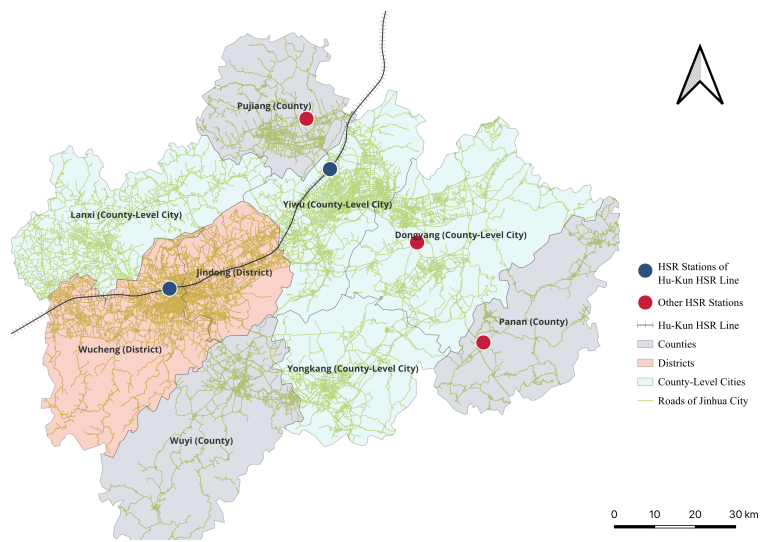

**Figure 1. High-speed rail routes and station locations in Jinhua City**

Table 1. Administrative Classification and Economic Profile of County-Level Divisions in Jinhua City

| Divisions | Name | Population (2020) | GDP (2023, 100 million RMB) | Area (km²) | Per Capita GDP (2023) |
|---|---|---|---|---|---|
| Urban District | Wucheng | 957,000 | 804.83 | 1,391.24 | 82500 |
|  | Jindong | 507,000 | 322.88 | 658.16 | 61900 |
| County-level City | Yiwu | 1,859,000 | 2,055.62 | 1,104.53 | 108400 |
|  | Dongyang | 1,088,000 | 805.85 | 1,746.81 | 73900 |
|  | Yongkang | 964,000 | 755.98 | 1,047.49 | 77600 |
|  | Lanxi | 575,000 | 496.75 | 1,312.44 | 85700 |
| County | Wuyi | 462,000 | 355.49 | 1,568.18 | 72500 |
|  | Pujiang | 461,000 | 291.46 | 918.16 | 62800 |
|  | Pan'an | 177,000 | 139.75 | 1,194.74 | 78300 |

## 3 Methods and Data

Understanding the economic consequences of HSR expansion requires a robust empirical strategy and a well-structured dataset that captures China's geographic and administrative diversity. This study focuses on two of China's most significant east–west HSR corridors: the Hu-Kun HSR (Shanghai–Kunming) and the Xu-Lan HSR (Xuzhou–Lanzhou) which together span vast regions and connect cities at varying stages of development across eastern, central, and western China. These lines are not only major components of the national HSR network but also serve as key arteries driving regional integration and transformation.

To rigorously evaluate the local impacts of HSR openings and the locations of HSR stations, we construct a balanced panel dataset covering the years 2008 to 2019, encompassing five years before and after the launch of HSR services along these corridors. After comprehensive data cleaning and validation, the final analytical sample consists of 353 county-level divisions within prefecture-level cities traversed by the selected HSR lines (Figure 2). Shanghai is excluded from the analysis because its unique status as a provincial-level municipality renders it fundamentally different from other samples.

Our identification strategy is grounded in the difference-in-differences (DID) approach, which is widely regarded as an effective method for assessing the causal effects of large-scale infrastructure investments (Sun et al., 2023; Yang et al., 2019; Jia et al., 2017; Zhang et al., 2022). The treatment group includes all county-level divisions within cities directly traversed by either the Hu-Kun or Xu-Lan corridors, while the control group is composed of county-level divisions in prefecture-level cities that did not gain HSR connections by 2019. This group structure ensures that our analysis captures both the direct and spillover effects of HSR expansion, while accounting for China's spatial and institutional heterogeneity.

The core outcome variables are the production values of the primary, secondary, and tertiary sectors, which we use as principal indicators of local economic development. The definitions and data sources of all variables used in the analysis are summarized in Table 2. By tracking changes in these sectoral outputs, we assess not only the overall impact of HSR on county-level economies but also the nuanced patterns of heterogeneity across regions and administrative types. Summary statistics, including the mean, standard deviation, minimum, and maximum values of sectoral output by event time, are presented in Table 3. Based on these variables and data, we now turn to the empirical analysis of how HSR openings have affected county-level industrial development.

To provide a comprehensive overview of the data sources, sample structure, and analytical workflow, the overall research design is depicted in Figure 3. This framework summarizes the major steps, including data construction, DID and event study analysis, subgroup and mechanism analysis, robustness checks, and the synthesis of results and policy implications.

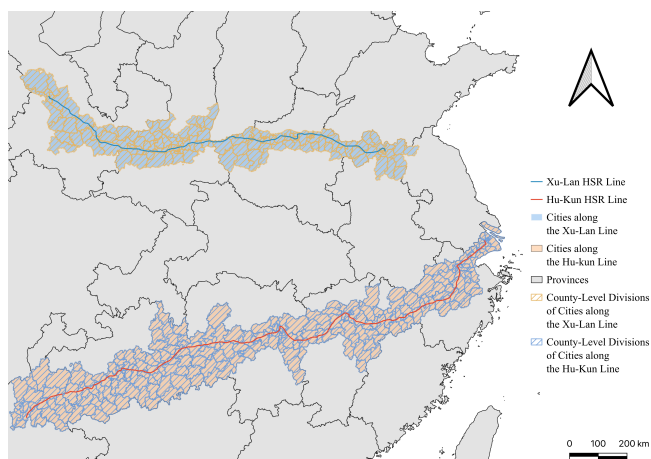

**Figure 2. Map of Hu-Kun and Xu-Lan HSR routes**

```
Sample: 353 County-level Divisions (Urban Districts, Counties, County-level Cities)
-> Panel Data (2008–2019)
-> Multiperoid DID Analysis (Overall & Subgroups: Region, Admin Type, HSR Station)
-> Event Study (Dynamic, Parallel Trends)
-> Mechanism Analysis (Population Mobility, Service Sector)
-> Robustness Checks (PSM-DID, Placebo)
-> Interaction Effects (e.g., HSR × Region, HSR × Admin Type, HSR × Station)
-> Main Results & Policy Recommendations
```

**Figure 3. Research design and analytical framework of this study**

**Table 2. Variable Definitions**

| Variable Name | Full Name | Description | Unit | Data Source |
|---|---|---|---|---|
| id | **County/City Identifier** | Unique code for each county-level administrative unit | | Local government statistical yearbooks |
| treat | **Treatment Group Indicator** | It equals one if the county/city receives an HSR opening and zero otherwise. | | Official railway opening records and local government reports |
| GDP1 | **Primary Sector Values Added** | Value-added of the primary sector (agriculture, forestry, animal husbandry, and fishery) | 100 million RMB | Local government |
| GDP2 | **Secondary Sector Value Added** | Value-added of the secondary sector (sector and construction) | 100 million RMB | Local government |
| GDP3 | **Tertiary Sector Values Added** | Value added by the tertiary sector (all service sectors) | 100 million RMB | Local government |
| Event Time | **Event Time Indicator** | Years relative to HSR opening for each county/city: -2, -1, 0, +1, +2 | Year | Constructed by the authors |

**Note:** All monetary values are measured at current prices. GDP data are compiled from local government statistical yearbooks and bulletins.

Table 3. Descriptive statistics of sectoral value-added by event time

| Variable | Event Time | N | Mean | Std. Dev. | Min | Max |
| --- | --- | --- | --- | --- | --- | --- |
| Primary Sector | −2 | 1143 | 16.35 | 14.99 | 0.00 | 95.26 |
| Secondary Sector | −2 | 1145 | 61.77 | 57.20 | 0.15 | 630.35 |
| Tertiary Sector | −2 | 1145 | 50.25 | 56.13 | 1.52 | 459.54 |
| Primary Sector | −1 | 1145 | 16.07 | 14.67 | 0.00 | 64.87 |
| Secondary Sector | −1 | 1145 | 61.70 | 56.53 | 0.08 | 583.97 |
| Tertiary Sector | −1 | 1145 | 53.63 | 59.10 | 1.91 | 467.09 |
| Primary Sector | 0 | 1145 | 15.88 | 14.54 | 0.00 | 64.87 |
| Secondary Sector | 0 | 1145 | 62.36 | 57.35 | 0.08 | 623.78 |
| Tertiary Sector | 0 | 1145 | 56.72 | 61.54 | 2.09 | 475.56 |
| Primary Sector | +1 | 1145 | 15.68 | 14.47 | 0.00 | 64.87 |
| Secondary Sector | +1 | 1145 | 62.36 | 57.34 | 0.08 | 623.78 |
| Tertiary Sector | +1 | 1145 | 59.74 | 64.25 | 2.29 | 495.86 |
| Primary Sector | +2 | 1145 | 15.53 | 14.48 | 0.00 | 64.87 |
| Secondary Sector | +2 | 1145 | 62.66 | 57.52 | 0.08 | 629.36 |
| Tertiary Sector | +2 | 1145 | 62.55 | 66.81 | 2.30 | 510.67 |

**Note:** negative values are years before opening, positive values are years after.

### 3.1 Basic Analysis of HSR's Impact on County-level Divisions

### 3.1.1 Overall Analysis

To provide a preliminary understanding, this section examines the effects of HSR implementation on the industrial structure of all county-level divisions in cities along the routes. The study also reveals how HSR influences the development of county-level divisions, offering important references for planning and implementing HSR projects. To formally evaluate this effect, we estimate the following DID model:

$$Y_{it} = \alpha + \beta Treat_i \times Post_t + \mu_i + \lambda_t + \epsilon_{it}$$

where $Y_{it}$ is the outcome for county or city $i$ in year $t$; $Treat_i$ is a treatment group indicator (one if HSR opened, zero otherwise); $Post_t$ is a posttreatment indicator; $\mu_i$ and $\lambda_t$ are county and year fixed effects; $\epsilon_{it}$ is the error term; $Region_i$ in 3.1.2 indicates the regional group (eastern, central, or western); $Division_i$ in 3.1.3 indicates the administrative type (urban district, county, or county-level city); $Station_i$ in 3.2.1 equals one if county or city $i$ has an HSR station and zero otherwise.

### 3.1.2 Analysis of Different Economic Zones

This section introduces the empirical strategy for assessing the effects of HSR on the industrial structure of county-level divisions across different economic zones. Based on China's economic and technological development status, the country is divided into three economic belts (Eastern, Central, and Western) which are all traversed by the two selected HSR corridors. Accordingly, all county-level divisions in the sample are classified into these three regions for analysis. To test for regional heterogeneity in the effects of HSR, the baseline model is extended by including interactions between the treatment variable and regional group indicators, as follows:

$$Y_{it} = \alpha + \beta_1 Treat_i \times Post_t + \beta_2 Treat_i \times Post_t \times Region_i + \mu_i + \lambda_t + \epsilon_{it}$$

### 3.1.3 Analysis of Different Administrative Divisions

This section describes the classification of county-level administrative divisions and the modeling approach used to examine the differential impacts of HSR. Few existing studies have specifically focused on county-level divisions, and even fewer have conducted in-depth analyses. Notably, China's county-level divisions differ substantially from those in other countries. In China, cities are organized as prefecture-level administrative units, with urban cores typically composed of *"urban districts"* within county-level administrative divisions. In contrast, *"counties"* and *"county-level cities"* are generally more independent and do not border the urban core. As a result, the effects of HSR may vary among "urban districts," "counties," and "county-level cities." Accordingly, this section examines the impact of HSR on the industrial structures across these different types of county-level divisions. To assess whether the impact of HSR varies by administrative type, we further interact the treatment with indicators for urban districts, counties, and county-level cities:

$$Y_{it} = \alpha + \beta_1 Treat_i \times Post_t + \beta_2 Treat_i \times Post_t \times Division_i + \mu_i + \lambda_t + \epsilon_{it}$$

### 3.2 Location Analysis of HSR Stations

#### 3.2.1 Overall Analysis

While most studies focus on the impact of HSR at the prefecture level, research rarely examines the effects of station location at the county level. Here, we analyze how the locations of HSR stations affect the economic development of county-level divisions. To evaluate how the presence of HSR stations influences local economic development, we estimate the following DID specification, incorporating an interaction between treatment and station location:

$$Y_{it} = \alpha + \beta_1 Treat_i \times Post_t + \beta_2 Treat_i \times Post_t \times Station_i + \mu_i + \lambda_t + \epsilon_{it}$$

#### 3.2.2 Analysis of Different Administrative Divisions

After analyzing the impact of station location on all county-level divisions, we hypothesize that the effects may differ among administrative divisions. To examine potential heterogeneity in the effects of station locations across administrative divisions, we extend the model by interacting the treatment, post-treatment, and station indicators with indicators for administrative type:

$$Y_{it} = \alpha + \beta_1 Treat_i \times Post_t + \beta_2 Treat_i \times Post_t \times Station_i + \beta_3 Treat_i \times Post_t \times Division_i \\ + \beta_4 Treat_i \times Post_t \times Station_i \times Division + \mu_i + \lambda_t + \epsilon_{it}$$

### 3.3 Event Study

To examine the dynamic effects of HSR openings and test the parallel trend assumption, we implement an event study. Specifically, we estimate the following model:

$$Y_{it} = \sum_{k \neq -1} \beta_k D_{i,t+k} + \alpha_i + \lambda_t + \epsilon_{it}$$

where $Y_{it}$ denotes the outcome variable for county $i$ at event time $t$, and $D_{i,t+k}$ is a set of dummy variables indicating years relative to the HSR opening (with $k = -1$ as the reference year). County and year fixed effects are also included. The coefficients $\beta_k$ trace the dynamic impacts before and after HSR openings. Parallel pre-trends are indicated by insignificant $\beta_k$ before HSR opening, but significant posttreatment coefficients reflect policy impacts.

### 3.4 Placebo Test

To ensure that spurious correlations or unobserved shocks do not drive the main results, we conduct a placebo test by randomly assigning the HSR opening ("treatment") status across counties and years, while retaining the dataset structure unchanged. Specifically, we repeat the random assignment and estimation process 1000 times, estimating the following model in each iteration:

$$Y_{it} = \alpha + \beta^{placebo} FakeTreat_i \times Post_t + \mu_i + \lambda_t + \epsilon_{it}$$

where $FakeTreat_i$ is a treatment indicator randomly assigned to mimic the true HSR opening. If the baseline effect is causal, the distribution of placebo coefficients ($\beta^{placebo}$) should center around zero and rarely approach the magnitude of the main result.

# 4 Results

This section presents empirical findings on the effects of HSR openings and station locations on county-level industrial development in China. The analysis begins by comparing aggregate economic outcomes before and after HSR implementation. Next, we examine how these changes vary across different regions (eastern, central, and western economic belts) as well as among distinct administrative divisions including urban districts, counties, and county-level cities. Finally, we assess the role of direct HSR station access in shaping these post-policy changes and highlight the resulting disparities across administrative units.

## 4.1 Impact of HSR Opening on County-Level Industrial Development

Table 4 reports the estimated effects of HSR openings on value-added across the three major sectors. The difference-in-differences estimation reveals that the opening of HSR lines resulted in a clear and statistically significant shift in the industrial development of county-level divisions along the Hu-Kun and Xu-Lan corridors. Comparing economic output before and after the HSR implementation, the most pronounced increases were observed in the secondary (manufacturing) and tertiary (services) sectors. The primary sector, which includes agriculture and resource-based activities, showed only a modest improvement after HSR opening. On average, the introduction of HSR increased value-added by approximately 42 million RMB in the primary sector, 224 million RMB in the secondary sector, and 220 million RMB in the tertiary sector.

These results suggest that HSR investments trigger substantial changes in local economic structures, with manufacturing and services experiencing the greatest acceleration after the policy intervention. The expansion of the secondary and tertiary sectors reflects the practical benefits of enhanced transport connectivity for the movement of people, goods, and information. For manufacturing, HSR contributes to lower logistics costs and stronger integration into national value chains, while for the service sector, greater mobility stimulates local demand and encourages knowledge exchange. In contrast, the limited gains in the primary sector indicate that agriculture and similar activities are less affected by improved intercity transport. To verify the validity and robustness of the DID estimation, a parallel trend test was conducted, as shown in Figure 4. This figure provides evidence that, prior to the opening of HSR, the outcome trends between the treatment and control groups were highly similar, thereby supporting the key identification assumption of the DID approach.

However, these average effects conceal significant differences across both regions and administrative divisions. The subsequent sections examine how these policy-driven changes unfold within different regional and administrative contexts, providing a more detailed understanding of the structural impact of HSR expansion. However, these average effects conceal significant differences across both regions and administrative divisions. The following sections provide a more detailed breakdown of these variations.

**Table 4. Opening Results for the Overall Analysis**

|  | Primary sector | Secondary sector | Tertiary sector |
| --- | --- | --- | --- |
| HSR opening | 0.42 | 2.24 | 2.20 |
|  | *** | *** | *** |
| P > \|t\| | 0.000 | 0.000 | 0.000 |
| N | 5,725 | 5,725 | 5,725 |

Note: *, **, and *** indicate significant levels at 10%, 5%, and 1%, respectively.

The value of industrial output is in RMB 100 million, as are all the tables below.

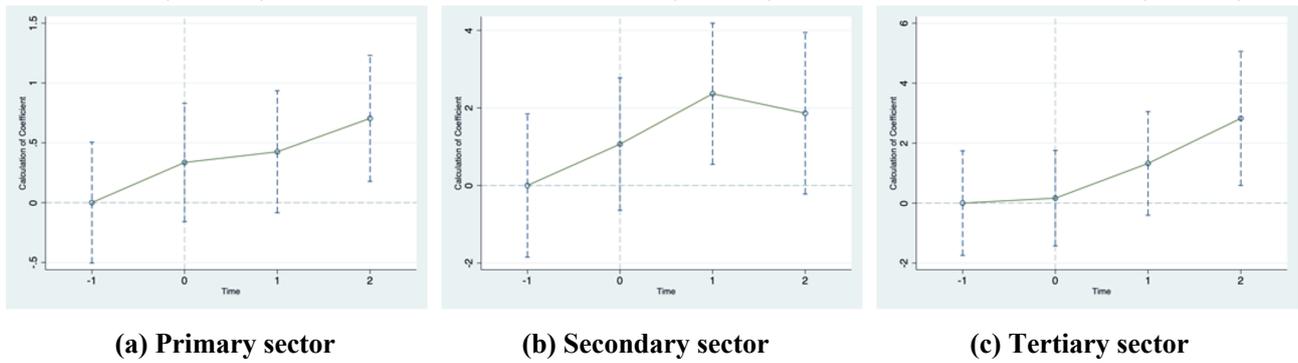

(a) Primary sector  (b) Secondary sector  (c) Tertiary sector

**Figure 4. Parallel trend test of openings for overall analysis**

## 4.2 Regional Heterogeneity: Eastern, Central, and Western China

Building on the overall findings, we next examine whether the effects of HSR expansion differ across major economic zones. Previous research has shown that HSR can significantly boost growth in major cities, while its impact on smaller or less-developed regions is often limited. Our analysis directly confirms these regional disparities by comparing economic outcomes before and after HSR openings across Eastern, Central, and Western China.

The results reveal that HSR openings generated highly uneven effects. In Eastern China, the most developed region, HSR introduction led to a substantial increase in the tertiary sector, with an average gain of nearly 1.29 billion RMB (Table 5), and strong growth in the secondary sector, reflecting large markets and mature industrial networks. These benefits are most concentrated in areas with advanced economic bases and dense service activity. In contrast, Central China saw only modest gains after HSR opening, with the secondary sector increasing by 144 million RMB and the tertiary sector by 192 million RMB, both much smaller than in the east. This pattern suggests that intermediate development levels and weaker integration may limit Central China's ability to capitalize on new infrastructure. Western China followed a different trajectory. Here, HSR expansion led to significant growth in the primary sector (102 million RMB) and the secondary sector (341 million RMB), likely due to the region's resource base, low initial industrialization, and the transformative impact of improved connectivity on previously isolated areas. The parallel trend test results for each region are shown in Figure 5.

These results highlight that the economic effects of HSR openings are far from uniform. Regional heterogeneity underscores the importance of local economic foundations and integration in shaping infrastructure outcomes. The benefits of HSR depend heavily on pre-existing conditions, and regions with weaker economic structures may require additional policy support to achieve similar gains.

**Table 5. Results of Opening for Different Economic Zones**

|  | Eastern China | | | Central China | | | Western China | | |
| --- | --- | --- | --- | --- | --- | --- | --- | --- | --- |
|  | Primary sector | Secondary sector | Tertiary sector | Primary sector | Secondary sector | Tertiary sector | Primary sector | Secondary sector | Tertiary sector |
| HSR opening | 0.42 *** | 0.75 | 12.91 *** | −0.25 | 1.44 * | 1.92 *** | 1.02 *** | 3.41 *** | −0.89 |
| P > \|t\| | 0.007 | 0.627 | 0.000 | 0.295 | 0.082 | 0.009 | 0.000 | 0.000 | 0.129 |
| N | 4,210 | 4,210 | 4,210 | 4,675 | 4,675 | 4,675 | 4,760 | 4,760 | 4,760 |

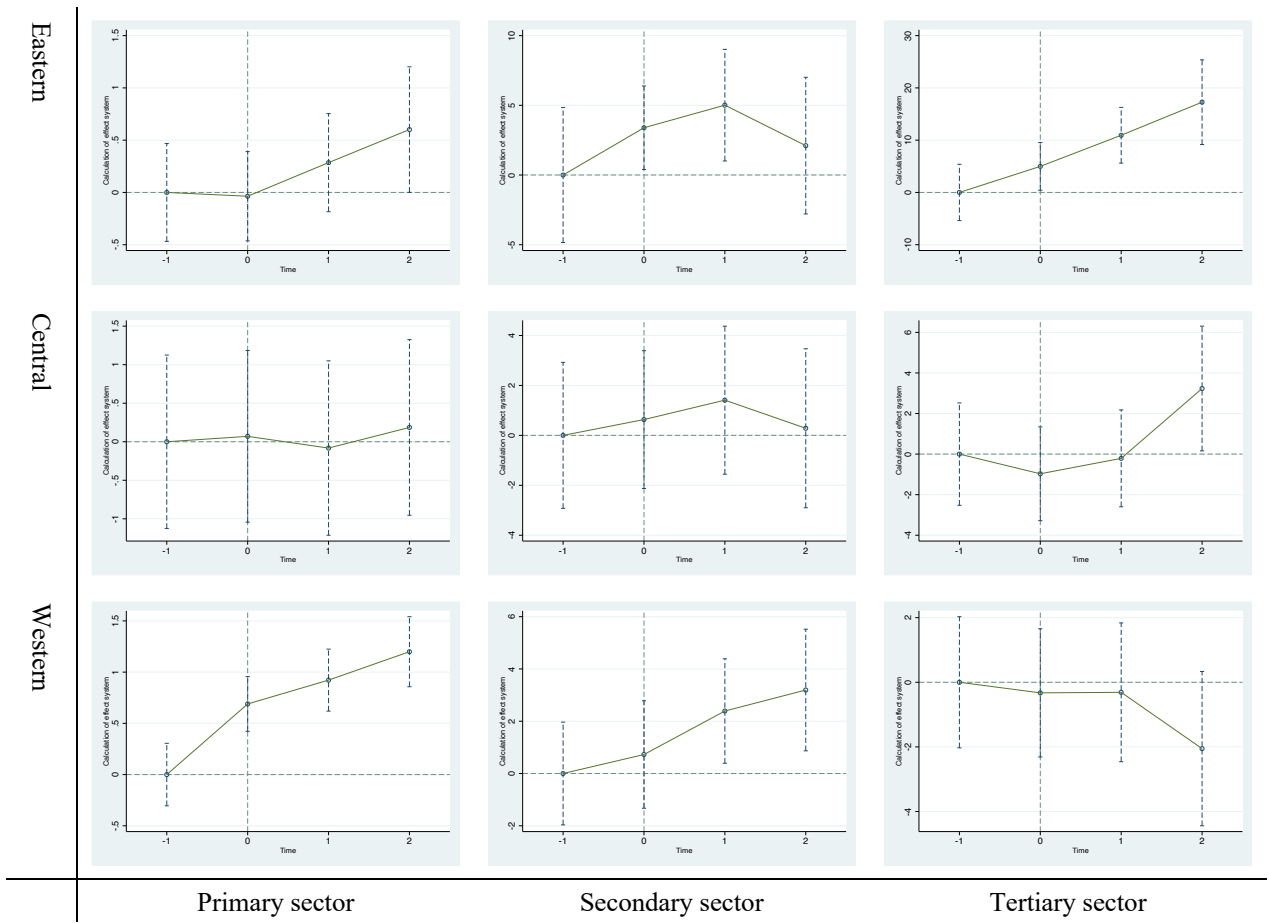

**Figure 5. Parallel trend test of openings for different economic zones**

## 4.3 Differences Among Administrative Divisions

Beyond regional disparities, the analysis also reveals substantial heterogeneity across administrative categories. The analysis also uncovers substantial differences among the various types of county-level divisions when comparing conditions before and after the HSR opening.

Urban districts, which are the most developed areas within prefecture-level cities, show the largest absolute gains after HSR implementation, especially in the service sector (Table 6). Districts benefit the most from improved mobility, and their strong population and business bases allow them to take advantage of HSR-driven opportunities. Manufacturing in urban districts also increases significantly following the policy intervention, further reflecting their capacity to integrate into wider supply chains and consumer markets. County-level cities, which possess both urban features and a certain level of administrative independence, also register significant improvements after HSR access. The secondary and tertiary sectors both expand, as improved transport connectivity attracts investment, skilled labor, and commercial activity.

In contrast, the impact of HSR openings on counties is distinctly different. While these areas observe some positive changes in manufacturing and the primary sector, the service sector experiences a notable contraction, declining by an average of 283 million RMB after HSR begins operation (Table 6). This negative shift suggests that HSR may accelerate the movement of people and resources out of less-developed counties, resulting in reduced local demand for services. Supplementary analysis using county-level population data confirms that HSR openings are associated with significant population outflows from counties, which in turn further weakens their service sector performance (Table 12). Figure 6 presents the parallel trend test by administrative division, confirming the robustness of the DID results across urban districts, counties, and county-level cities.

Table 6. Results of Openings for Different Administrative Divisions

| | Urban Districts | | | Counties | | | County-Level Cities | | |
|---|---|---|---|---|---|---|---|---|---|
| | Primary sector | Secondary sector | Tertiary sector | Primary sector | Secondary sector | Tertiary sector | Primary sector | Secondary sector | Tertiary sector |
| HSR opening | 0.28 | 2.77 ** | 8.71 *** | 0.38 ** | 1.08 ** | −2.83 *** | 0.45 * | 5.48 *** | 5.54 *** |
| P > \|t\| | 0.295 | 0.018 | 0.000 | 0.040 | 0.012 | 0.000 | 0.066 | 0.000 | 0.002 |
| N | 4,555 | 4,555 | 4,555 | 4,900 | 4,900 | 4,900 | 4,200 | 4,200 | 4,200 |

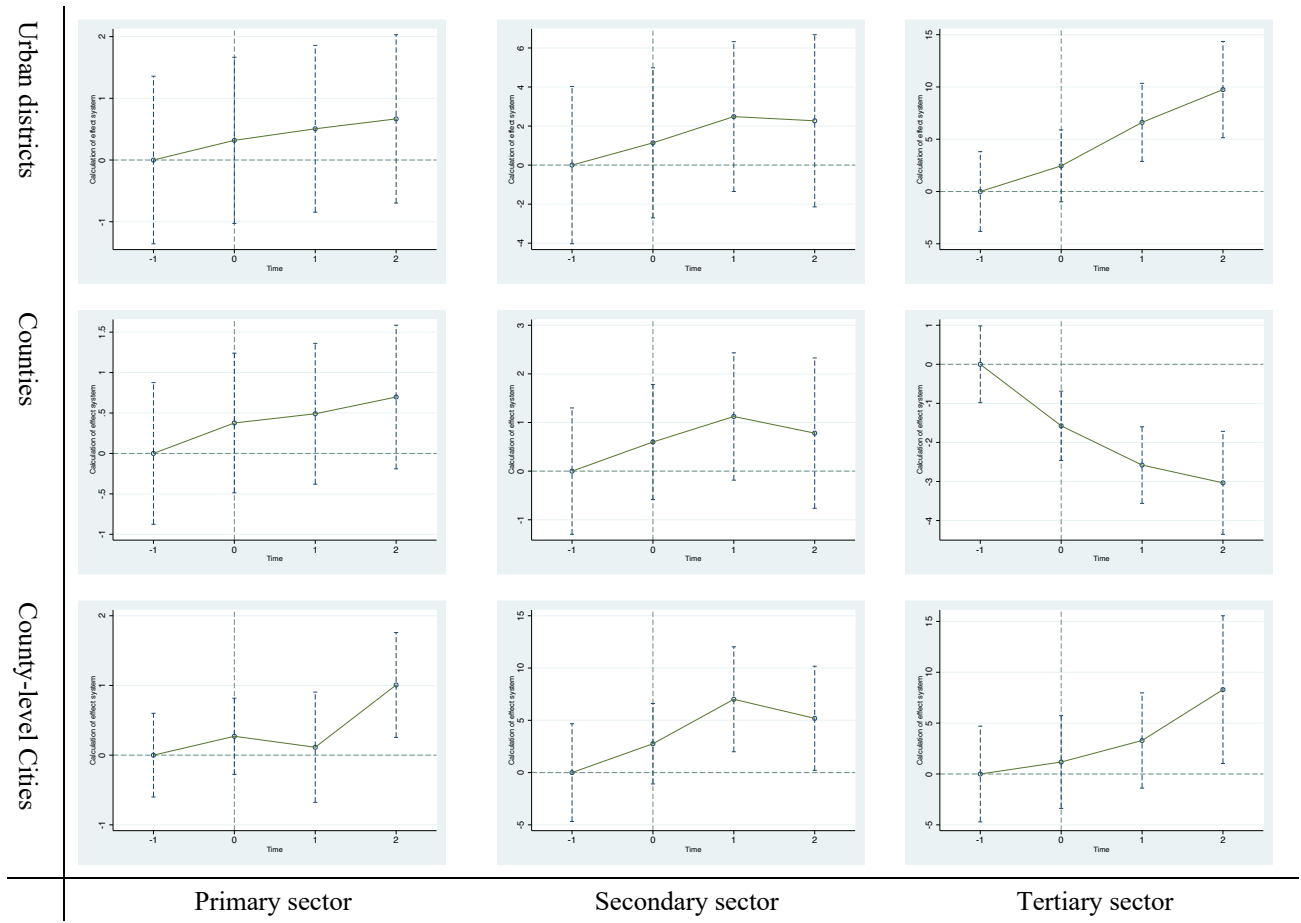

Figure 6. Parallel trend test of openings for different administrative divisions

## 4.4 HSR-Stations Locations' Amplifying Role

To further clarify the source of these disparities, we explicitly assess how direct access to HSR stations affects local economic outcomes. The results indicate that county-level divisions with direct HSR stations experience much greater economic gains after HSR implementation compared to those without stations. In counties and urban districts with station access, the secondary sector increased by an average of 431 million RMB, and the tertiary sector by 631 million RMB (Table 7). In contrast, for counties without HSR stations, the increases were only 175 million RMB and 124 million RMB.

These findings show that the benefits of HSR are concentrated in locations with immediate access to stations, supporting the idea that direct connectivity is essential for unlocking agglomeration effects. While limited positive spillovers are observed in areas without stations, the scale of these effects is much smaller, which highlights the importance of station placement in determining the overall impact of HSR expansion. Figure 7 shows the parallel trend test by HSR station access.

**Table 7. Results of Stations for Overall Analysis**

|  | With HSR Stations | | | Without HSR stations, | | |
|---|---|---|---|---|---|---|
|  | Primary sector | Secondary sector | Tertiary sector | Primary sector | Secondary sector | Tertiary sector |
| HSR stations*Opening | −0.07 | 4.31 *** | 6.31 *** | 0.53 *** | 1.75 *** | 1.24 ** |
| P > \|t\| | 0.889 | 0.003 | 0.000 | 0.000 | 0.002 | 0.025 |
| N | 4,295 | 4,295 | 4,295 | 5,390 | 5,390 | 5,390 |

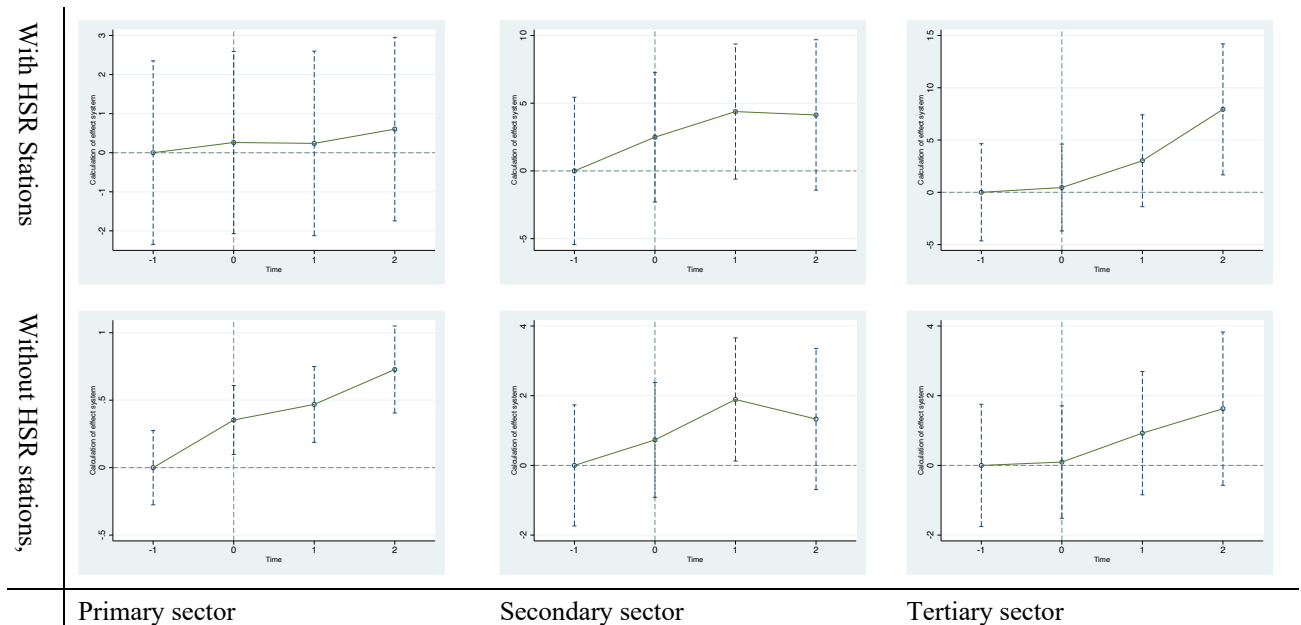

Figure 7. Parallel trend test of stations for overall analysis

## 4.5 Stations Effects Across Administrative Divisions

Extending the analysis to the interplay between administrative divisions and station access, our results show that the benefits of HSR openings are not distributed equally. This section reports the differentiated impacts of HSR station locations across urban districts, counties, and county-level cities (Table 8–10). Figures 8–10 present the parallel trend tests by administrative division and station access. Urban districts, as developed urban cores, benefit most from direct HSR station access, with significant gains in the tertiary sector and notable increases in manufacturing output. These findings support the view that agglomeration economies and service-sector growth are most easily activated in densely populated, economically advanced areas. County-level cities, which combine urban features with administrative flexibility, experience strong growth in both the secondary and tertiary sectors when equipped with an HSR station. This underscores the role of improved connectivity in positioning these cities as new investment and labor hubs within the HSR network.

For counties, the effects are more nuanced. Having an HSR station can partially mitigate negative trends, resulting in smaller contractions in the tertiary sector and modest gains in manufacturing compared to counties without stations. However, many counties still experience population and business outflows to better-developed areas. Previous research has shown that HSR can constrain counties' secondary sector development by reducing fixed asset investment (Yu et al., 2021), and our results further suggest that simply having a station is often insufficient to reverse these trends. This pattern indicates that improved accessibility alone is not enough; complementary interventions such as business incentives, skills training, and better local amenities are needed for less-developed counties to fully benefit from major infrastructure investments.

Overall, these results demonstrate that both administrative structure and direct HSR station access together determine the extent and direction of local economic impacts, which aligns with broader research (Yu et al., 2021). These findings highlight the importance of strategic HSR station placement and targeted support for less-advantaged counties.

**Table 8. Results of Stations in Urban Districts**

|  | With HSR Stations ||| Without HSR stations, |||
| --- | --- | --- | --- | --- | --- | --- |
|  | Primary sector | Secondary sector | Tertiary sector | Primary sector | Secondary sector | Tertiary sector |
| HSR Stations*Opening | −0.71 | 7.27 *** | 10.57 *** | 0.48 *** | 1.64 | 8.55 *** |
| P > \|t\| | 0.514 | 0.009 | 0.000 | 0.000 | 0.175 | 0.000 |
| N | 4,874 | 4,874 | 4,874 | 4,404 | 4,404 | 4,404 |

**Table 9. Results of Stations in Counties**

|  | With HSR Stations ||| Without HSR stations, |||
| --- | --- | --- | --- | --- | --- | --- |
|  | Primary sector | Secondary sector | Tertiary sector | Primary sector | Secondary sector | Tertiary sector |
| HSR Stations*Opening | −1.27 | 0.60 | −1.65 ** | 0.58 *** | 1.11 ** | −2.99 *** |
| P > \|t\| | 0.326 | 0.433 | 0.020 | 0.000 | 0.011 | 0.000 |
| N | 4,075 | 4,075 | 4,075 | 4,780 | 4,780 | 4,780 |

**Table 10. Results of Stations in County-Level Cities**

|  | With HSR Stations ||| Without HSR stations, |||
| --- | --- | --- | --- | --- | --- | --- |
|  | Primary sector | Secondary sector | Tertiary sector | Primary sector | Secondary sector | Tertiary sector |
| HSR Stations*Opening | 0.48 | 6.08 ** | 12.11 ** | 0.44 * | 5.21 *** | 2.55 ** |
| P > \|t\| | 0.401 | 0.041 | 0.016 | 0.067 | 0.003 | 0.040 |
| N | 4,035 | 4,035 | 4,035 | 4,125 | 4,125 | 4,125 |

|With HSR Stations| 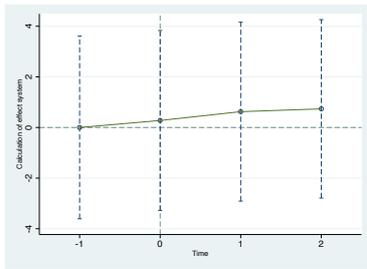 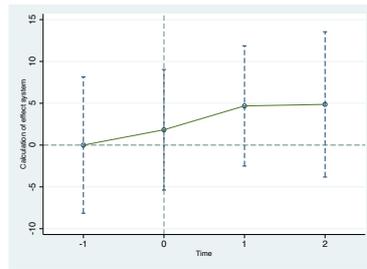 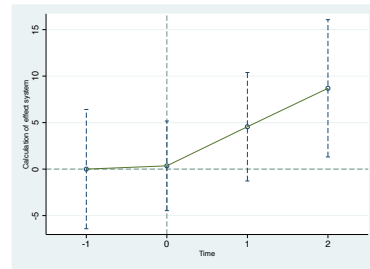
|---|---|
|Without HSR stations,| 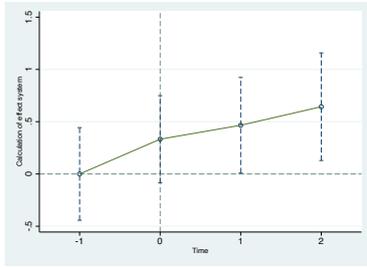 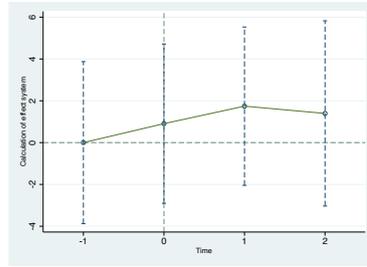 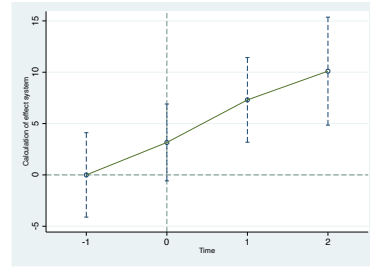

Primary Sector　　　　　　Secondary Sector　　　　　　Tertiary sector

**Figure 8. Parallel trend test of stations in urban districts**

|With HSR Stations| 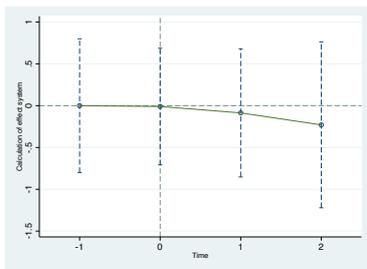 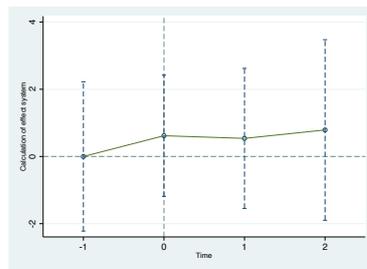 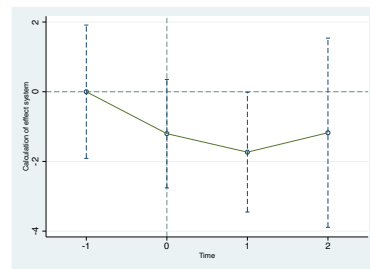
|---|---|
|Without HSR stations,| 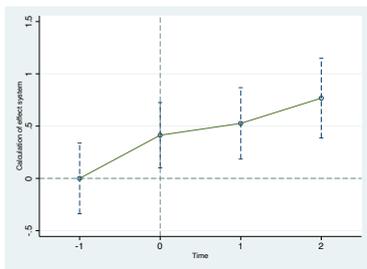 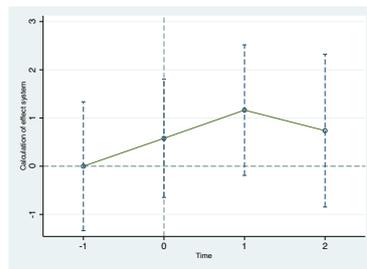 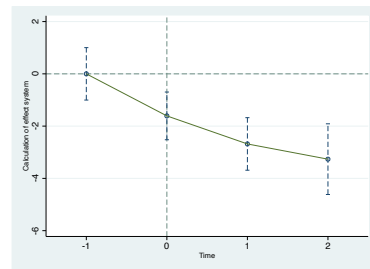

Primary Sector　　　　　　Secondary Sector　　　　　　Tertiary Sector

**Figure 9. Parallel trend test of stations in counties**

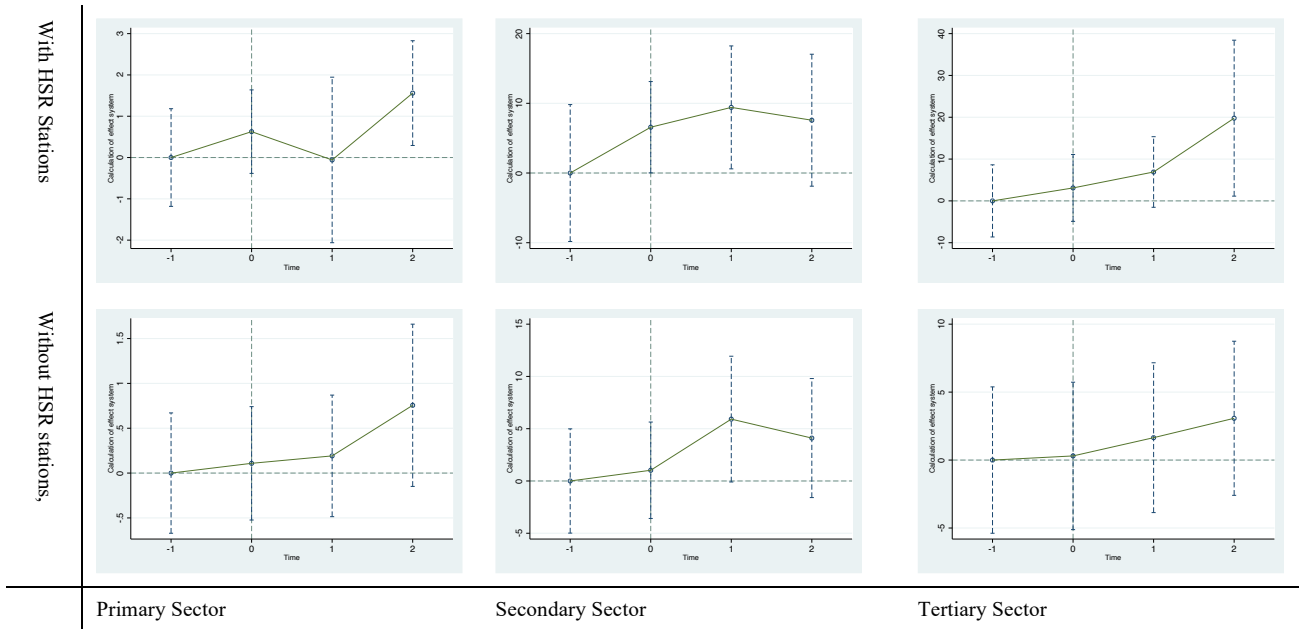

**Figure 10. Parallel trend test of stations in county-level cities**

## 4.6 Event Study

To further assess the robustness of our findings and to capture the dynamic impact of HSR policy, we conduct an event-study analysis based on Table 11. The results show that during the two years before the introduction of HSR, there are no significant changes in the output of any sector. This absence of pre-existing trends supports the parallel trends assumption and further validates the reliability of our DID framework.

Following the HSR opening, a clear structural shift emerges. In the two years after policy implementation, the primary sector exhibits a sustained and statistically significant decline, indicating a contraction in agricultural output. By contrast, both the secondary and, especially, the tertiary sectors display marked and persistent growth, with the service sector showing the strongest upward trend. These results demonstrate that HSR openings serve as a turning point for county-level economic structure, accelerating industrial upgrading and service sector expansion while reducing the relative importance of agriculture. The consistency of these findings with our main regression results underscores the robustness and credibility of our empirical strategy.

**Table 11. Event-Study Results**

| Event Time | Primary Sector | 95% CI | Secondary Sector | 95% CI | Tertiary Sector | 95% CI |
|---|---|---|---|---|---|---|
| −2 | 0.25 | [0.06, 0.43] | 0.07 | [−0.56, 0.69] | −3.38 | [−3.88, −2.89] |
| 0 | −0.19 | [−0.27, −0.11] | 0.66 | [0.07, 1.24] | 3.08 | [2.65, 3.52] |
| +1 | −0.40 | [−0.53, −0.26] | 1.46 | [0.52, 2.40] | 6.08 | [5.34, 6.81] |
| +2 | −0.49 | [−0.66, −0.33] | 0.91 | [−0.27, 2.09] | 8.90 | [7.84, 9.96] |

Note: The estimates are from an event-study specification with event time 1 as the reference period. The county fixed effects are included, and standard errors are clustered at the county level. "Primary/Secondary/Tertiary Sector" refers to value added by the respective sectors. The coefficients represent differences relative to event time 1.

## 4.7 Placebo Test

To further assess the robustness and causal validity of our main findings, we conduct a comprehensive placebo test, following established practice in policy evaluation literature (Li et al., 2016). The central aim of this exercise is to determine whether the observed treatment effects could plausibly be driven by spurious correlations, unobserved shocks, or model specification errors, rather than by the actual implementation of HSR.

Specifically, we randomly reassign the HSR opening years and locations across the sample counties, while preserving the overall panel structure and all other variables. This procedure effectively breaks the genuine link between HSR openings and county-level economic outcomes, simulating a scenario in which the timing and placement of HSR are entirely unrelated to local economic trends. For each randomized iteration, we re-estimate the baseline DID model and record the resulting coefficients for the pseudo-treatment effect. This process is repeated 1000 times, generating a distribution of placebo estimates under the null hypothesis that HSR openings have no real impact.

The underlying logic is as follows: if our main results merely reflect chance correlations or omitted variable bias, then the coefficients derived from these randomized experiments should frequently attain values like or greater than those observed in the actual analysis. Conversely, if the true effect of HSR is substantive, then the placebo distribution should be tightly centered around zero, with the observed estimate from the real data lying well outside this distribution.

Figure 11 illustrates the results of the placebo test. The histogram displays the distribution of placebo coefficients obtained from the 1000 randomized simulations, while the red vertical line indicates the actual treatment effect estimated in our primary analysis. It is clear that the vast majority of placebo estimates cluster closely around zero, and none approach the magnitude of the true effect. This pronounced divergence provides compelling evidence that our main findings are not artifacts of random chance, unobserved shocks, or model misspecification, but rather reflect a genuine causal impact of HSR openings on local economic development.

In summary, the placebo test reinforces the credibility and robustness of our identification strategy. By demonstrating that random assignment of HSR openings fails to generate effects of comparable size or significance, we strengthen the claim that the observed treatment effects in our study are indeed attributable to the actual introduction of HSR.

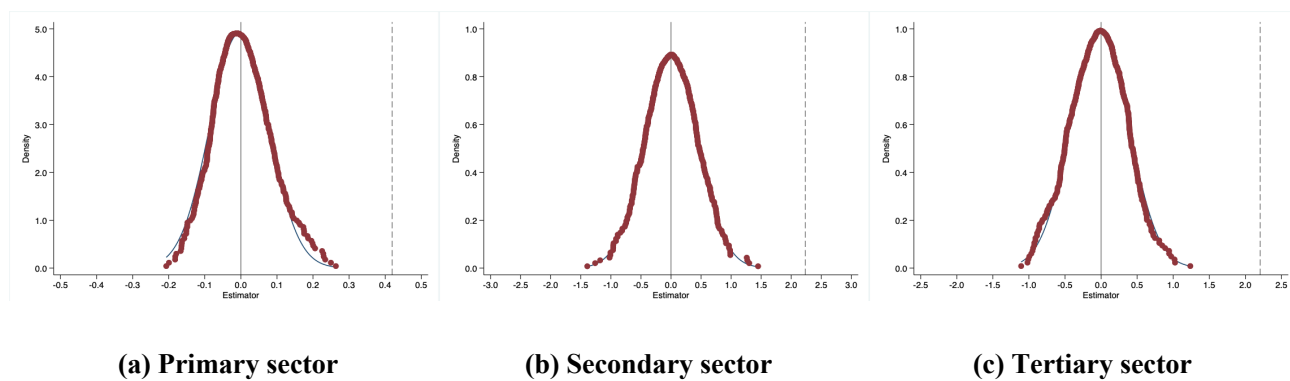

(a) Primary sector  (b) Secondary sector  (c) Tertiary sector

**Figure 11. Placebo test**

## 4.8 Summary of Key Findings

Bringing together the above analyses, our results clearly demonstrate that the opening of HSR lines serves as a significant policy turning point for county-level economies in China. The most notable improvements occur in manufacturing and service sectors, particularly in regions and divisions that possess strong economic foundations and have direct access to HSR stations. In the two years following HSR implementation, these areas achieve substantial and sustained growth, confirming the transformative effect of improved connectivity.

However, this policy impact is far from uniform. Counties without HSR stations or with weaker development bases not only fail to share equally in the benefits, but in some cases experience declining service sector activity and increasing population outflows after HSR introduction. This pronounced divergence highlights that HSR policy, while stimulating growth in some localities, can intensify regional and administrative disparities if implemented without accompanying targeted interventions. These findings underscore the necessity of integrating large-scale infrastructure investment with locally adapted policy support to ensure that the economic gains from HSR expansion are balanced, inclusive, and sustainable.

# 5 Discussion

## 5.1 Mechanism Analysis of Population Outflow's Role

Building on the results above, we further investigate why HSR openings may suppress service sector development in counties by conducting a supplementary DID analysis using counties' population data (Table 12). The estimated log-linear coefficient (−0.324, p < 0.01) indicates that counties affected by HSR experience an average population decline of about 28 percent. This substantial outflow suggests that improved connectivity accelerates the movement of residents out of these areas, which in turn reduces local demand for services and contributes to the contraction of the tertiary sector observed after the policy intervention.

This mechanism-focused analysis extends beyond the aggregate findings of recent studies such as Ou et al. (2025), who documented significant urban–rural disparities and spillover effects of HSR at the county level but primarily relied on economic output indicators. While Ou et al. highlighted spatial heterogeneity in HSR's impacts, their analysis did not directly examine the role of demographic change or explicitly test how population outflows shape sectoral shifts. By empirically linking population decline to the weakening of the service sector, our findings provide new evidence on the causal pathways through which HSR alters local economic structures. This approach helps clarify not only the distributional consequences of HSR expansion but also the mechanisms that generate uneven development across regions.

However, it is important to note that China's county-level population statistics are subject to considerable limitations, including missing values, inconsistencies in statistical definitions, and possible local reporting biases. As such, this population-based analysis should be seen as supplementary support for the underlying mechanism rather than definitive causal proof. The primary conclusions of this study remain grounded in the more reliable economic output indicators presented in the main analysis.

In addition to population outflow, there may be other mechanisms contributing to the heterogeneous effects of HSR expansion on county-level economies. For example, changes in local public investment, adjustments in business location decisions, and shifts in local industrial or fiscal policies may also influence economic outcomes after HSR implementation. Due to the constraints of currently available data, this study is unable to empirically test these additional channels. Nevertheless, the possibility of their influence cannot be excluded. Future research that incorporates firm-level information, local government expenditure data, or survey-based measures of business and household expectations could provide a more comprehensive understanding of the multiple pathways through which HSR affects local economic structures.

**Table 12. Effect of HSR on Counties' Population**

| Variable | Coefficient | Std. Error | t-value | P > \|t\| | 95% Confidence Interval | |
|---|---|---|---|---|---|---|
| Log_Population | −0.324*** | 0.029 | −11.32 | 0.000 | −0.380 | −0.268 |
| Constant | 3.834*** | 0.022 | 175.37 | 0.000 | 3.791 | 3.877 |
| Notes: Robust standard errors are reported. | | | | | | |

## 5.2 Robustness Check: Pre-GDP Propensity Score Matching

To further strengthen the credibility of our main results, we conducted a propensity score matching (PSM) analysis using average pre-HSR GDP as the matching variable. This approach substantially improved the comparability between treated and control county-level divisions by ensuring a balanced baseline in terms of economic development. As presented in Table 13 and Figure 12, the estimated effects of HSR openings on industrial output in the matched sample are highly consistent with our main analysis: significant and sustained increases in both the secondary and, most notably, the tertiary sectors, while the primary sector remains largely unchanged. These findings confirm that the observed policy effects are not simply the result of pre-existing differences in economic scale, thereby reinforcing the robustness and reliability of our core conclusions.

Table 13. Estimated Effects of HSR on County-Level Divisions (PSM-DID)

|  | Primary Sector | Secondary Sector | Tertiary Sector |
| --- | --- | --- | --- |
| DID | 0.28* (0.15) [0.062] | 1.08 (0.68) [0.111] | 2.64*** (0.69) [0.000] |
| Year 1 | −0.23 (0.23) [0.325] | 1.64* (0.89) [0.066] | 4.36*** (0.91) [0.000] |
| Year 0 | 0.01 (0.23) [0.967] | 2.10** (0.89) [0.019] | 6.93*** (0.91) [0.000] |
| Year 1 | 0.08 (0.23) [0.730] | 2.96*** (0.89) [0.001] | 9.76*** (0.91) [0.000] |
| Year 2 | 0.33 (0.23) [0.152] | 4.06*** (0.89) [0.000] | 11.64*** (0.91) [0.000] |
| Observations | 3,033 | 3,035 | 3,035 |
| Adjusted R² | 0.976 | 0.983 | 0.983 |

Notes: Robust standard errors appear in parentheses. All models include city and year fixed effects.

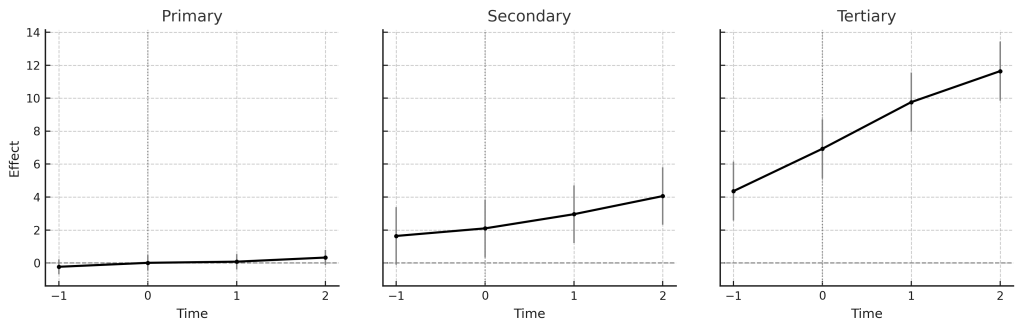

Figure 12. Parallel trend test (PSM-DID)

## 5.3 Discussion of the Basic Analysis

Building on the results in Sections 4.2 and 4.3, as well as the preceding robustness analyses, we further explore how the economic effects of HSR openings differ across both regions and administrative divisions. As shown in Figure 13, the benefits of HSR expansion remain highly uneven, shaped by regional context and local industrial structure. In the two years following HSR implementation, the tertiary sector in Eastern China experiences the most pronounced growth, with an average increase of nearly 1.3 billion RMB (12.91 × 100 million RMB). This finding highlights that service-oriented regions with advanced market structures and dense populations benefit most from HSR, as enhanced connectivity promotes agglomeration and market access.

The secondary sector also sees substantial gains, particularly in the west, where the increase reaches approximately 341 million RMB (3.41 × 100 million RMB). This suggests that improved transport connectivity can have transformative effects even in less developed areas. In contrast, Central China's gains are relatively modest in all sectors, reflecting its intermediate economic base.

The right panel of Figure 13 reveals marked disparities across different types of county-level divisions. Urban districts and county-level cities exhibit strong growth in both the secondary and tertiary sectors after HSR opening, which can be attributed to agglomeration effects, supply chain integration, and increased mobility. In stark contrast, counties experience a significant decline in the tertiary sector, averaging a loss of 283 million RMB (−2.83 × 100 million RMB). This points to a risk of "economic hollowing" in peripheral regions. The likely causes are resource siphoning and population outflows toward more developed urban centers as a result of HSR (Qin, 2017; Hall, 2009). As discussed in Section 5.1, our supplementary analysis using county-level population data suggests that this decline in services may be partly driven by accelerated population outflows following HSR implementation, which reduce local demand for services.

However, it is important to acknowledge that other factors, such as changes in local investment or shifts in industrial policy, may also contribute to the observed contraction in the tertiary sector. The primary sector remains largely unaffected across all regions and administrative types.

Taken together, these findings underscore that large-scale transport infrastructure alone is insufficient to achieve balanced regional development. While HSR can drive growth in regions and divisions with strong economic foundations and direct access, it may also intensify spatial and administrative disparities, particularly for less-developed counties without stations. Achieving inclusive growth thus requires not only infrastructure investment but also the implementation of locally adapted policies, such as targeted human capital investment, business support, and enhanced regional linkages. These measures are essential to help peripheral counties benefit from HSR and avoid decline. In summary, the spatial disparities revealed in this county-level analysis highlight the need to integrate HSR expansion with broader development strategies and targeted policy interventions, ensuring the transformative potential of HSR is realized equitably across China's diverse regions.

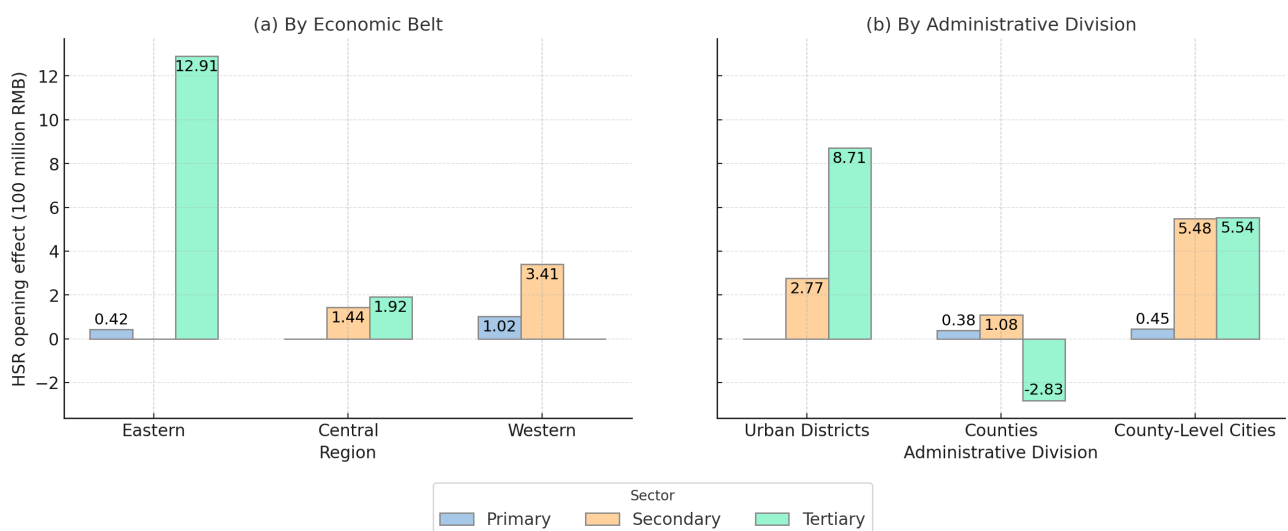

Figure 13. Results for HSR openings' impact

## 5.4 Discussion for Analysis of HSR Stations' Locations

Extending the analysis of policy effects to the spatial dimension, we find that the location of HSR stations is a critical determinant of economic impact at the county level, especially when comparing periods before and after HSR implementation. As demonstrated in Figure 14 and in line with the differentiated impacts observed in earlier sections, urban districts and county-level cities with direct HSR access experience the most pronounced gains in both manufacturing and service sectors following the introduction of HSR. For example, county-level cities with HSR stations saw tertiary sector growth of 1.211 billion RMB and secondary sector gains of 608 million RMB, while urban districts with HSR stations recorded 1.057 billion RMB and 727 million RMB in these sectors, respectively. These benefits stem from reduced transportation costs, enhanced market connectivity, and greater ability to attract skilled labor and investment, mechanisms that are well documented in international research (Willigers & Van Wee, 2011; Murakami & Cervero, 2017).

However, our findings also indicate that the presence of an HSR station does not guarantee uniformly positive outcomes, particularly for less-developed counties. Counties without HSR stations experienced a significant contraction in the tertiary sector, averaging a loss of 299 million RMB, and even those with stations saw only limited gains. Persistent challenges, including limited economic scale and weak demographic dynamics, continue to constrain the transformative potential of improved accessibility in these areas. The observed divergence between counties and county-level cities after HSR openings further emphasizes that infrastructure investment alone is insufficient to stimulate sustained growth in less-developed counties.

The economic impact of HSR stations thus depends heavily on local capacity, administrative flexibility, and the presence of complementary policy interventions. County-level cities, which generally have larger populations and more diversified economic bases, are better able to leverage HSR access for long-term development. In contrast, peripheral counties, even those with stations, may continue to lag behind after HSR implementation. This underscores the importance of integrated planning, in which HSR network expansion is coordinated with targeted economic support, investment in feeder transport, and policies designed to address regional disparities. Previous studies on major rail infrastructure projects have emphasized the need to consider spatial externalities and market imperfections in cost-benefit analyses (Elhorst & Oosterhaven, 2008). It should also be noted that this study is limited by the accuracy of spatial data and the potential influence of unobserved local policies. Future research should address these issues by using more detailed data and employing advanced identification strategies (such as improved matching methods or natural experiments) to better isolate causal effects.

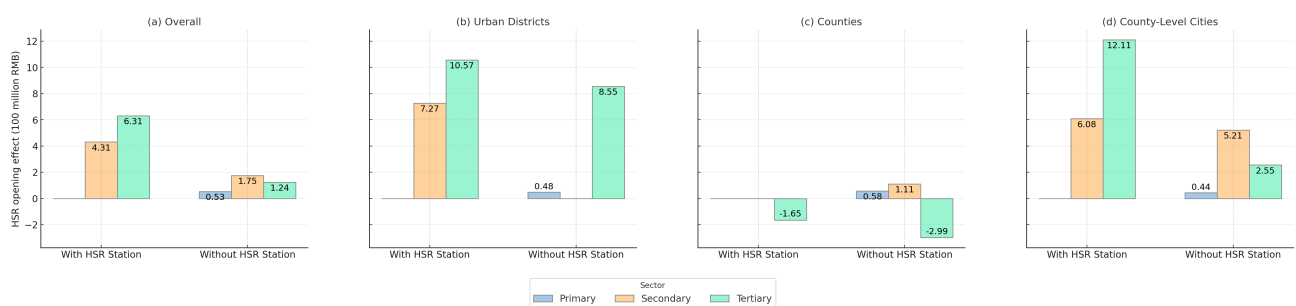

**Figure 14. Results for HSR stations' impact**

# 6 Conclusion

Synthesizing the findings from the previous sections, this study demonstrates that the expansion of HSR has brought about significant, yet highly uneven, economic changes across China's county-level divisions. The evidence presented in Sections 4 and 5 consistently shows that, after HSR openings, most county-level divisions along the rail routes experienced substantial increases in value added from both secondary and tertiary sectors, with the most pronounced gains concentrated in urban districts and county-level cities. For instance, in the eastern region, county-level divisions with direct HSR stations recorded average growth in the tertiary sector exceeding 1.2 billion RMB, while both the eastern and western regions saw increases of over 300 million RMB in the secondary sector. These outcomes highlight that HSR expansion can serve as a catalyst for local economic upgrading, particularly in areas with robust market foundations and strong regional integration.

However, the advantages brought by HSR are far from universal. As discussed in Sections 4.4 and 5.4, county-level divisions with direct station access generally realized more significant improvements in both manufacturing and services, whereas those without stations experienced only modest changes or, in some cases, no clear growth at all. Moreover, regional variation remains a key feature of these impacts. The eastern region achieved the largest and most sustained growth following HSR expansion, while the western region, despite being less developed, benefited from rapid initial increases, likely reflecting greater marginal returns to connectivity. In contrast, Central China, with its intermediate level of economic development, responded less strongly and more slowly to HSR, underscoring the critical role of pre-existing conditions.

Importantly, as revealed in both the event-study analysis and the supplementary findings in Section 5.1, many less-developed county-level divisions experienced a decline in service sector output after HSR implementation, a trend closely associated with accelerated population outflows. This raises the concern that, without appropriate policy interventions, HSR expansion may intensify the concentration of resources and population in already advantaged areas and deepen regional disparities.

Overall, the impact of HSR on local development is multifaceted and strongly dependent on local circumstances. The results make clear that investment in infrastructure alone is not sufficient to ensure balanced regional growth. Future policy efforts must address the underlying inequalities in access and local capacity by complementing infrastructure investment with targeted support, such as improving human capital, enhancing feeder transport systems, and fostering inclusive economic environments. Only through such integrated approaches can the full benefits of HSR expansion be realized across all regions.

## 7 Policy Implications

The analyses in Sections 4 and 5 demonstrate that the expansion of HSR has significantly reshaped economic development patterns across China's county-level divisions, with clear differences emerging between regions and administrative types. Urban districts and county-level cities with direct HSR stations have achieved sustained growth in manufacturing and services, reflecting the connectivity and agglomeration advantages identified in both previous research and this study. In contrast, many counties, particularly those lacking direct HSR access, have experienced contractions in the service sector and increased population outflows, as shown by the mechanism analysis in Section 5.1. These divergent trajectories highlight the need for policy responses that are closely tailored to local and regional conditions.

Our findings are broadly consistent with the recent work of Ou et al. (2025), which emphasize the widening urban–rural divide and the presence of spatial spillover effects in HSR development. However, this study advances the policy discussion by revealing how differences in station accessibility and administrative status within county-level divisions can generate even greater heterogeneity in development outcomes. Effective policy design must therefore move beyond general resource allocation and address the specific institutional and spatial challenges faced by different types of localities.

A regionally and administratively differentiated policy framework is essential to maximize the benefits of HSR while minimizing spatial disparities. Figure 15 summarizes this logic. The process begins with the identification of regional and administrative heterogeneity, followed by the design and implementation of targeted policy measures for each locality type in eastern, central, and western China.

- In the eastern region, where urban districts and county-level cities have seen the greatest gains, policy should focus on further promoting innovation, supporting the growth of high-end service clusters near HSR stations, and enhancing urban integration. Additional investments in public amenities and local transport can help consolidate these advantages and generate positive spillovers to neighboring areas.
- In the central region, policies should encourage industrial upgrading and supply chain extension, especially in urban districts and county-level cities, and improve last-mile connectivity. Supporting interregional cooperation can also help to leverage the opportunities brought by HSR expansion.
- In the western region, priorities should include developing regional specialties and green industries, upgrading basic infrastructure, and increasing fiscal and talent support, particularly in less-developed counties.
- For counties, especially those without direct HSR access, policy interventions should address the risks of economic hollowing and service sector decline. This can be achieved by strengthening local industry support (including manufacturing and agriculture), providing business incentives and improved feeder transport, and implementing measures to retain and attract human capital. Enhancing public services such as healthcare, education, and community amenities can help stabilize population and support local consumption. Where necessary, additional fiscal transfers and targeted assistance from higher-level governments may be needed to maintain essential services and economic stability.
- Across all regions, continuous monitoring, evaluation, and policy adjustment are crucial to ensure that interventions remain effective and responsive to changing local circumstances.

This regionally tailored, adaptive approach is essential for achieving inclusive and balanced regional development in the HSR era. By closely coordinating infrastructure investment with targeted policy measures, governments can ensure that the economic benefits of HSR expansion reach all regions and communities, rather than exacerbating existing disparities.

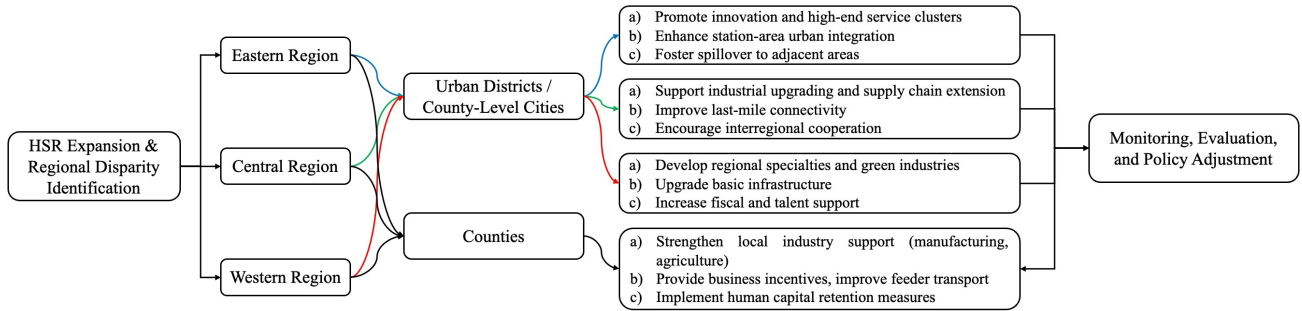

**Figure 15. Framework for Regionally Differentiated Policy Responses to HSR Expansion**

# 8 Research Perspectives

While this study provides a thorough assessment of the impacts of HSR on county-level divisions, several important limitations should be acknowledged. Most of these limitations are a result of current constraints in data availability in China. For example, the lack of detailed information on household income, enterprise activities, and precise patterns of population movement makes it difficult to fully explain the mechanisms behind the observed changes or to control for all potential confounding factors.

Another significant limitation concerns the analysis of HSR station placement. The location of HSR stations is not randomly assigned, but is often influenced by local economic, political, and demographic factors (Ahlfeldt & Feddersen, 2018; Willigers & Van Wee, 2011). Because relevant data such as project planning documents, local lobbying records, or detailed land-use information are not available, this study could not implement more advanced identification strategies such as instrumental variables or natural experiments to separate the causal effect of station placement from other local characteristics. As a result, although strong associations between direct station access and economic outcomes are observed, it is possible that some of these effects may be driven by unobserved factors that influence both station location and local economic performance.

In addition, the relatively short observation period after HSR opening means that it is not possible to assess longer-term effects, especially in regions where structural changes may take more time to appear. The mechanism analysis in this study mainly relies on aggregate data such as population outflows and does not directly examine alternative channels, including shifts in public investment, business relocation, or changes in local policies, due to data limitations.

These challenges are not unique to this research. Recent studies, such as Ou et al. (2025), have noted similar difficulties in causal identification and data completeness when working at the county level. Both studies face the reality that data constraints limit the ability to disentangle all relevant mechanisms or fully isolate the independent impact of HSR investment decisions. However, the present analysis sought to address these challenges by making full use of available data, conducting a series of robustness checks, and supplementing the main results with descriptive and mechanism-based analyses focusing on population change and economic output. These efforts enhance the credibility of the findings, but they cannot fully overcome the limitations caused by missing or incomplete data and constraints on causal inference.

As data collection and access continue to improve, future research will be able to use more detailed spatial and temporal data, longer time periods, and a wider range of control variables to provide deeper insights, especially regarding the mechanisms of impact and the causal effects of station placement. Integrating firm-level data, household surveys, and high-frequency spatial indicators such as satellite imagery would enable researchers to more precisely identify the channels through which HSR affects local economies.

Despite these limitations, the methods and strategies used in this study represent an important step forward in understanding the complex effects of HSR on regional development in China. Continued progress in regional data systems and more detailed future research will be essential for deepening this field of inquiry and supporting more equitable and effective infrastructure policy.